\documentclass[pre,letterpaper,floatfix,superscriptaddress,11pt]{revtex4}
\usepackage{graphicx}

\begin{document}

\title{Local Simulation Algorithms for Coulomb Gases with Dynamical Dielectric Effects} 

\author{A.~Duncan}
\affiliation{Department of Physics and Astronomy\\
University of Pittsburgh, Pittsburgh, PA 15260}
\author{R.D.~Sedgewick}
\author{R.D.~Coalson}
\affiliation{Department of Chemistry\\
University of Pittsburgh, Pittsburgh, PA 15260}


\begin{abstract}

We discuss the application of the local lattice technique of Maggs and
Rossetto \cite{maggs:prl} to problems that involve the motion of
objects with different dielectric constants than the background.  In
these systems the simulation method produces a spurious interaction
force which causes the particles to move in an unphysical manner.  We
show that this term can be removed using a variant of a method known
from high-energy physics simulations, the multiboson method, and
demonstrate the effectiveness of this corrective method on a system of
neutral particles.  We then apply our method to a one-component plasma
to show the effect of the spurious interaction term on a charged
system.

\end{abstract}
\maketitle

\section{Introduction}

The long-range character of the electrostatic Coulomb interaction lies
at the root of the computational difficulties encountered in the
simulation of many systems of biophysical interest, in which one
wishes to understand the thermodynamics of a Coulomb gas of
charged particles --- or macroions --- moving in an environment of
spatially varying dielectric. Recently, Maggs and collaborators
\cite{maggs:prl,maggs:alone,maggs:worm, maggs:trail} have suggested
rewriting the problem of a classical Coulomb gas in a local lattice
framework in which each charged particle responds only to the local
electric field, which is also updated so that Gauss' Law is
respected at each point in the simulation. Most applications of the
original local algorithm, as well as a variety of suggested
improvements \cite{ourpaper,ourpaper:fft}, have dealt primarily with
situations in which the dielectric constant is not spatially
varying. In this case, the transverse (or ``curl") part of the
electric field, which is unconstrained in the partition sum,
decouples from the physics of the charged ions, so that the method
correctly calculates the classical electrostatic energy of the
charges.
  
In this paper, we consider situations in which the dielectric constant
becomes dynamical: i.e. is spatially varying, and in a way that
depends on the location of the charges in the system, so that it
changes in the course of the Monte Carlo simulation. In a molecular
dynamics simulation, for example, one effectively determines the
electrostatic energy of the system at each step by solving the
appropriate Poisson equation (in the presence of the varying
dielectric).  Of course, the resulting energy expression still suffers
from the problem of having to include contributions from all pairs of
charged particles in the system.  In the local lattice approach, on
the other hand, the unconstrained transverse part of the functional
integral over the electric field no longer decouples from the charged
particle dynamics, and leads \cite{maggs:alone} to an apparent
attractive dipole-dipole interaction, even when the system is
completely electrically neutral, with zero classical
electrostatic energy. Although such interactions certainly exist at a
quantum level (e.g. Casimir, Keesom, van der Waals forces), our
interest here is in a purely classical calculation.
    
Our purpose in this paper is to show that the local algorithm of Maggs
et al can be extended by addition of a set of boson fields --- with a
completely local hamiltonian --- which remove any unphysical
contributions from the curl part of the electric field, so that the
thermodynamics reflects the classical partition function of a charged
system with electrostatic energy $H_{\rm es}=\int
d\vec{r}\frac{\vec{D}^{2}}{2\epsilon(\vec{r})}$, as desired. In
Section~\ref{sec:theproblem} we describe in detail the structure of
the transverse field contribution in a local lattice formulation of
the Coulomb gas. In particular, we derive the form of the determinant
induced by this contribution. In Section~\ref{sec:howitworks}, we
describe the use of a multiboson technique familiar in lattice gauge
theory (``unquenched" quantum chromodynamics) to eliminate the
spurious determinant factor. In Section~\ref{sec:results}, we test the
method by calculating the density-density structure factor, both with
a system of neutral dielectric particles and with a one-component
charged plasma.  Finally, in Section~\ref{sec:conclusion}, we
summarize our results and discuss the outlook for further
applications.
    
\section{Transverse Mode Contributions in Systems with Varying Dielectric}
\label{sec:theproblem}

In this section we shall show how the local algorithm of references
\cite{maggs:prl, maggs:alone,maggs:worm,
maggs:trail,ourpaper,ourpaper:fft} can be modified to prevent the
appearance of spurious dipole-like interactions that are not present
in the classical electrostatics of the system when the dielectric
constant varies both spatially and in the course of the simulation
(e.g. if there are mobile entities with dielectric different from that
of the ambient medium). We illustrate the point first in the case of a
continuous system. Later, a spatial lattice will be introduced to make
the functional integrations precise.


Consider the partition function of a system consisting of $N$ free
charges (mobile or fixed) $e_i$ at locations $\vec{r}_{i}$, thereby
producing a free charge density
\begin{equation}
\label{eq:chargedens}
   \rho(\vec{r}) = \sum_{i}e_{i}\delta(\vec{r}-\vec{r}_{i})
\end{equation}
where the system is also described by a linear dielectric function
(here assumed isotropic) $\epsilon(\vec{r})$.  Note that in general the
dielectric function $\epsilon(\vec{r})$ may depend on the locations of
the free charges $\vec{r}_{i}$, which may be embedded in regions of
varying dielectric. This dependence should be kept in mind, although
for notational convenience it will be suppressed in the following.
The electric displacement can be broken into longitudinal and
transverse parts using the
general Helmholtz decomposition,
\begin{eqnarray}
\label{eq:helmdecomp1}
  \vec{D}(\vec{r}) &=& -\epsilon(\vec{r})\vec{\nabla}\phi(\vec{r})+\vec{\nabla}\times\vec{A}(\vec{r}) \\
  \label{eq:helmdecomp2}
                 &=& \vec{D}^{||}(\vec{r})+\vec{D}^{\rm tr}(\vec{r}).
\end{eqnarray}
As the transverse and longitudinal components are orthogonal to each
other, we have
\begin{equation}
\label{eq:decouple}
  \int d\vec{r}\frac{\vec{D}^{2}}{\epsilon(\vec{r})}=\int d\vec{r}\frac{\vec{D}^{||}(\vec{r})^{2}}{\epsilon(\vec{r})}
  +\int d\vec{r}\frac{\vec{D}^{\rm tr}(\vec{r})^{2}}{\epsilon(\vec{r})}.
  \end{equation}

If the constraint of Maxwell's second law is explicitly imposed
the transverse part of the electric displacement vanishes and the
electrostatic energy of the system is given by
\begin{equation}
\label{eq:eenergy}
 H_{\rm es} = \frac{1}{2}\int d\vec{r}\, \frac{\vec{D}^{||}(\vec{r})^{2}}{\epsilon(\vec{r})}
 \end{equation}
and the   canonical partition function for the system at
inverse temperature $\beta$ becomes
\begin{equation}
\label{eq:partfunc}
  Z=\int\prod_{i=1}^{N}d\vec{r}_{i}e^{-\beta H_{\rm es}}
\end{equation}
where $\vec{D}^{||}$ must be determined by first solving
$-\vec{\nabla}\cdot(\epsilon\vec{\nabla}\phi)=\rho$, from which one
obtains
$\vec{D}^{||}=-\epsilon(\vec{r})\vec{\nabla}\phi(\vec{r})$. Note that
the (nonexistent) transverse part of the displacement field plays no
role in this result.
  
On the other hand, the partition function proposed in
\cite{maggs:alone} includes an integral over both the transverse and
longitudinal parts of the electric displacement and reads simply
\begin{eqnarray}
\label{eq:maggspart1}
    Z^{\prime}&=&\int\prod_{i=1}^{N}d\vec{r}_{i}\prod_{\vec{r}}{\cal
      D}\vec{D}(\vec{r})\,\delta(\vec{\nabla}
\cdot\vec{D}-\rho(\vec{r}))\,e^{-\frac{\beta}{2}\int d\vec{r}\vec{D}^{2}/\epsilon
(\vec{r})} \\
   &=&\int\prod_{i=1}^{N}d\vec{r}_{i}\prod_{\vec{r}}{\cal D}\vec{D}^{||}(\vec{r})\,{\cal D}\vec{D}^{\rm tr}(\vec{r})
   \,\delta(\vec{\nabla}\cdot\vec{D^{||}}-\rho(\vec{r}))   \nonumber \\
   \label{eq:maggspart2}
   &&\times e^{-\frac{\beta}{2}\int d\vec{r}\vec{D}^{||}(\vec{r})^{2}/\epsilon
(\vec{r})}e^{-\frac{\beta}{2}\int d\vec{r}\vec{D}^{\rm tr}(\vec{r})^{2}/\epsilon(\vec{r})} 
\end{eqnarray}
as a result of the identity Eq.\ \ref{eq:decouple}. It is apparent
that the integration over transverse degrees of freedom $\vec{D}^{\rm
tr}$ in Eq.\ \ref{eq:maggspart2} necessarily introduces a dependence
on the particle locations $\vec{r}_{i}$ through the dependence on
$\epsilon(\vec{r})$ absent in the conventional electrostatic energy of
Eq. \ref{eq:eenergy}. The exact form of this spurious
$\epsilon$-dependence can be uncovered by considering the form of
$Z^{\prime}$ for the special case $e_{i}=0$ of uncharged particles,
where $D^{||}=0$ and the entire dependence on $\epsilon(\vec{r})$
derives from the transverse part of the functional integral in
$Z^{\prime}$. 

Of course, in the real world quantum fluctuations of the transverse
(and longitudinal) parts of the field do exist, and would have to be
included in a fully quantum-mechanical treatment of the Coulomb gas
problem. This is not, however, the problem being addressed here, where
we are computing the purely classical partition function of a
classical Coulomb gas. Accordingly, the spurious interaction
potentials induced by the integration over transverse degrees of
freedom in Eq.\ \ref{eq:maggspart2} should not be identified with
Keesom potentials, for example, which have a quantum mechanical
origin, and a strength dependent on Planck's constant, which appears
nowhere in our classical partition function. Such non-Coulombic
potentials can of course be included phenomenologically in our
classical treatment as an explicit separate contribution to the energy
function (with an appropriate interaction strength).  For example,
hydrophobic interactions, which depend on intermolecular forces and
thus are quantum mechanical in origin, play a significant role in the
organization and function of molecular level biophysical structures
(e.g., membranes, proteins and nucleic acids).  Hydrophobic effects
can be included empirically by adding to the electrostatic free energy
a term proportional (with constant of proportionality often called the
``surface tension'') to the surface area of the membrane, protein, or
nucleic acid that is exposed to water \cite{sitkoff,kessel}.

Before investigating the special case of neutral particles, we shall
go over to a lattice discretization of the system. We imagine a
lattice Coulomb gas with displacement vector field $D_{n\mu}$ defined
on lattice links, where lattice sites are denoted $n$ and $\mu=1,2,3$
indicating the spatial direction of the link. Likewise, it turns out
to be convenient to associate a dielectric value with each {\em link}
(rather than site), so that the dielectric function on the lattice
becomes $\epsilon_{n\mu}$.  For example, in problems involving
macroions extending over several lattice sites and with dielectric
constant differing from that of the ambient medium, links crossing the
surface of the macroion can be chosen to have an appropriately
interpolated value of the dielectric constant. Alternatively, this can
be regarded as a generalization to allow nonisotropic systems where
the principal axes of the dielectric tensor coincide with the lattice
axes. The entire discussion given below can readily be generalized to
the case of a completely general dielectric tensor field. Introducing
left (resp. right) lattice derivatives $\bar{\Delta}$
(resp. $\Delta$), the Helmholtz decomposition of the displacement
field on the lattice becomes
\begin{equation}
\label{eq:helmlat}
  D_{n\mu} = -\epsilon_{n\mu}\Delta_{\mu}\phi_{n} + \sum_{\nu \sigma} \epsilon_{\mu\nu\sigma}\bar{\Delta}_{\nu}A_{n\sigma}
\end{equation}
where the electrostatic potential (lattice site field) $\phi_{n}$ satisfies the Poisson equation
\begin{eqnarray}
\label{eq:lattPoiss}
 -\sum_{\mu}\bar{\Delta}_{\mu}(\epsilon_{n\mu}\Delta_{\mu}\phi_{n}) &=& \rho_{n} \\
\label{eq:lattdens}
 \rho_{n} &=& \sum_{i} e_{i}\delta_{nr_{i}}
 \end{eqnarray}
In the absence of free charges, $e_i=0$, the local form of the partition function $Z^{\prime}$ becomes
\begin{equation}
\label{eq:Fdefine1}
 Z^{\prime}(e_{i}=0) \equiv \int \prod_{i=1}^{N}d\vec{r}_{i}{\cal F}(\epsilon)
\end{equation}
where the $\epsilon$-dependence is entirely due to the transverse
degrees of freedom and enters through the function
\begin{eqnarray}
\label{eq:Fdefine2}
 {\cal F}(\epsilon) &=& \int\prod_{n\mu}dD_{n\mu}\,\delta(\sum_{\mu}\bar{\Delta}_{\mu}D_{n\mu})\,
 e^{-\frac{\beta}{2}\sum_{n\mu}D_{n\mu}^{2}/\epsilon_{n\mu} } \\
\label{eq:Fdefine3}
 &=&\int\prod_{n}d\lambda_{n}\prod_{n\mu}dD_{n\mu}e^{i\sum_{n}\lambda_{n}\bar{\Delta}_{\mu}D_{n\mu}
  -\frac{\beta}{2}\sum_{n}D_{n\mu}^{2}/\epsilon_{n\mu}} \\
\label{eq:Fdefine4}
 &=&\int\prod_{n}d\lambda_{n}\prod_{n\mu}dD_{n\mu}\,e^{-i\sum_{n\mu}D_{n\mu}\Delta_{\mu}\lambda_{n}
  -\frac{\beta}{2}\sum_{n\mu}D_{n\mu}^{2}/\epsilon_{n\mu}} \\
\label{eq:Fdefine5}
 &=& C\prod_{n\mu}\sqrt{\epsilon_{n\mu}}\int \prod_{n}d\lambda_{n}\,e^{-\frac{1}{2\beta}\sum_{n\mu}\epsilon_{n\mu}
 (\Delta_{\mu}\lambda_{n})^{2}} \\
 \label{eq:Fdefine6}
 &=& C^{\prime}\prod_{n\mu}\sqrt{\epsilon_{n\mu}}\,{\rm det}^{-\frac{1}{2}}(-\sum_{\mu}\bar{\Delta}_{\mu}\epsilon_{\mu}\Delta_{\mu})
 \end{eqnarray}
In going from (\ref{eq:Fdefine2}) to (\ref{eq:Fdefine3}) we have
introduced an auxiliary field $\lambda_{n}$ to implement the Gauss'
Law constraint (for an everywhere neutral system), allowing the
Gaussian integration over the displacement field $D_{n\mu}$ to be
carried out explicitly. The constants $C$ and $C^{\prime}$ are independent
of $\epsilon$ and of no further interest. The integration over the
auxiliary $\lambda_{n}$ field (also Gaussian!)  then yields the
determinant of the indicated operator, whose action on a lattice site
field takes the explicit form
\begin{equation}
 \label{eq:defop}
  {\cal
    M}\lambda_{n}\equiv(-\sum_{\mu}\bar{\Delta}_{\mu}\epsilon_{\mu}\Delta_{\mu})\lambda_{n}
=\sum_{i=1}^{6}\epsilon_{ni}\lambda_{n}-\sum_{i=1}^{6}
  \epsilon_{ni}\lambda_{n+i}
  \end{equation}
where the index $i$ now runs over both positive ($i=1,2,3$) and
negative ($i=4,5,6$) spatial directions, and
$\epsilon_{ni}\equiv\epsilon_{n+i,i-3}$ for $i=4,5,6$.

In order to generate an ensemble of configurations based on the
partition function Eq. \ref{eq:partfunc} arising from the purely
physical electrostatic energy, the spurious functional dependence of
${\cal F}(\epsilon)$ must be removed: in other words, we should
insert a factor
\begin{equation}
  \label{eq:invF}
   {\cal F}^{-1}(\epsilon) = e^{-\frac{1}{2}\sum_{n\mu}\log{(\epsilon_{n\mu}})}{\rm det}^{+\frac{1}{2}}({\cal M})
   \end{equation}
into the partition integral (\ref{eq:maggspart1}) to remove the effect
of the transverse field modes. In the event that the dielectric
function is truly spatially-independent, or is spatially varying but
frozen throughout the simulation, adding the factor of ${\cal
F}^{-1}(\epsilon)$ is unnecessary, as the transverse integration
decouples from the dynamics of the problem.  

The problem that we are faced with here is well known: positive
(fractional or integral) powers of determinants of local operators are
intrinsically nonlocal, unlike negative powers, which may be
reexpressed as integrals over auxiliary fields with local actions
(cf. Eqs. (\ref{eq:Fdefine2}) through (\ref{eq:Fdefine6})). In lattice
quantum chromodynamics (QCD), for example, the inclusion of virtual
quark-antiquark processes leads to precisely the positive power of the
determinant of the Dirac operator in the path integral for the system,
greatly increasing the difficulty of simulations in the full
(``unquenched") version of the theory in comparison to the truncated
(``quenched") version where the quark determinant is simply dropped
\cite{QCD:unquenched}.
  
In the next section we shall show how the multiboson technique
introduced by L\"uscher (\cite{luescher}) for approximately computing
the positive power of quark determinants can be used to write a
\textit{purely local} Hamiltonian in terms of a set of auxiliary
scalars (``multibosons" in the lattice QCD language) which is readily
susceptible to Monte Carlo simulation and will allow us to implement
the correct electrostatic partition function (\ref{eq:partfunc}) for
systems with general dielectric makeup.

\section{Elimination of Transverse Contributions using Multiboson Fields}
\label{sec:howitworks}

The representation of determinants of local operators by integrals
over multiple auxiliary scalar fields was first introduced by L\"uscher
\cite{luescher} in the context of unquenched lattice QCD.  The
essential point is to find a uniform polynominal approximation to the
function $1/s$ in the interval [$\delta,1$] for small $\delta$. In
terms of the complex roots of the polynominal $z_k=\mu_k+i\nu_k$, a
convenient choice \cite{luescher} is the Chebyshev polynominal of order
$2N_B$ with
\begin{eqnarray}
 \label{eq:chebpoly}
 \frac{1}{s} &\simeq&  P(s) \equiv  C\prod_{k=1}^{N_B}((s-\mu_{k})^{2}+\nu_{k}^{2})  \\
  \mu_{k} &=& \frac{1}{2}(1+\delta)(1-\cos{\frac{2\pi k}{2N_B+1}}) \\
  \nu_{k} &=& \sqrt{\delta}\sin{\frac{2\pi k}{2N_B+1}} 
\end{eqnarray}
The representation of Eq.\ \ref{eq:chebpoly} can be extended to the
determinant of a real positive symmetric operator with spectrum in the
interval [$0,1$]
\begin{equation}
\label{eq:detpoly}
  {\rm det}^{+\frac{1}{2}}({\cal M}) \simeq \prod_{k=1}^{N_B} {\rm det}^{-\frac{1}{2}} (({\cal M}-\mu_{k})^{2}+\nu_{k}^{2})
\end{equation}
where the representation becomes exact in the limit
$N_B\rightarrow\infty,\delta\rightarrow 0$. However, we shall see that
accurate results can be obtained with surprisingly small values of $N_B$
(=\# of auxiliary scalar fields, see below).
 
The essential idea of the multiboson approach is to replace the
determinant factors on the right of Eq.\ \ref{eq:detpoly} by
integrals over a set of auxiliary fields $\phi^{(k)}_{n},\;k=1,2,..N_B$,
with \textit{local actions}, as follows
\begin{equation}
 \label{eq:multibos}
  \prod_{k=1}^{N_B} {\rm det}^{-\frac{1}{2}} (({\cal M}-\mu_{k})^{2}+\nu_{k}^{2})=C\int\prod_{k=1}^{N_B}
  {\cal D}\phi^{(k)}e^{-\sum_{k=1}^{N_B}\phi^{(k)}(({\cal M}-\mu_{k})^{2}+\nu_{k}^{2})\phi^{(k)}}
\end{equation}
where, once again, $C$ is an irrelevant constant that can henceforth
be neglected. As pointed out previously, our polynominal representation
assumes that the spectrum of the operator ${\cal M}$ in
Eq.\ \ref{eq:multibos} lies in the interval [$0,1$]. Let us assume
that the dielectric function $\epsilon$ of our system is bounded above
by the value $\epsilon_0$ (frequently, in biophysical simulations,
this is $\simeq$80, corresponding to the dielectric constant of an
aqueous ambient medium). Recalling that the lattice Laplacian operator
has largest eigenvalue equal to 12, one easily shows that the operator
\begin{equation}
\label{eq:defop2}
  {\cal M}\equiv \frac{1}{K\epsilon_0}(-\bar{\Delta}_{\mu}\epsilon_{\mu}\Delta_{\mu})
\end{equation}
has a spectrum contained in the unit interval provided $K\geq 12$. In
the multiboson approach, the auxiliary scalar field $\phi^{(k)}$ is
entrusted with resolving the spectrum of the operator ${\cal M}$ in
the neighborhood of $\mu_k$ in a region of width $\nu_k$ (see
\cite{luescher}).  For $k$ near 1 or $N_B$, $\nu_k$ is small and
$\mu_k$ is close to zero or one, and only a small region of the
spectrum is accurately treated. We can eliminate the lack of
resolution for finite $N_B$ at the upper end of the spectrum by
choosing $K$ somewhat larger than 12 (in the simulations reported
below, we typically take $K$=13), but if the spectrum of ${\cal M}$ is
very dense near the origin, we will necessarily be forced to use a
large value of $N_B$.  Fortunately, in the systems we have so far
simulated, this does not appear to be the case. This is a fortunate
distinction from the case of lattice quantum chromodynamics, where
chiral symmetry breaking necessarily implies a dense spectrum of
eigenvalues of the quark Dirac operator at the origin!  Instead, in
the systems of concern to us, the region with large dielectric
$\epsilon_0$ occupies almost all of the volume of the system, and the
spectral density of ${\cal M}$ does not differ appreciably from that
of the free Laplacian.
  
To summarize our proposal, we shall consider a system of mobile
charged entities (charges $e_i$, locations $\vec{r}_{i}$) with
total Hamiltonian
\begin{equation}
 \label{eq:htot}
   H_{\rm tot} = H_{\rm es}+ V_{\rm nc}(\vec{r}_{i})
\end{equation}
where the purely electrostatic energy $H_{\rm es}$ is given in
Eq. (\ref{eq:eenergy}), while all other non-Coulombic energy effects
(exclusion, Keesom, van der Waals, hard-core, soft-core, etc. etc.)
are included in $V_{\rm nc}$, with appropriate phenomenological
values.  The latter are not typically problematic as they are
essentially short-range effects. From
Eqs. (\ref{eq:Fdefine6},\ref{eq:invF},\ref{eq:multibos}) we find that
the correct expression for the partition function is
\begin{eqnarray}
   Z&=&
   \int\prod_{i}d\vec{r}_{i}\prod_{n\mu}dD_{n\mu}\,\delta(\sum_{\mu}
   \bar{\Delta}_{\mu}D_{n\mu}-\rho_{n})\nonumber \\
   \label{eq:partfinal1}
   &&\times
   {\cal F}^{-1}(\epsilon)\,e^{-\beta V_{\rm nc}}e^{-\frac{\beta}{2}\sum_{n\mu}D_{n\mu}^{2}/\epsilon_{n\mu}} \\
   &=&
   \int\prod_{i}d\vec{r}_{i}\prod_{n\mu}dD_{n\mu}\prod_{kn}d\phi^{(k)}_{n}
    \,\delta(\sum_{\mu}\bar{\Delta}_{\mu}D_{n\mu}-\rho_{n})\nonumber \\
  &&\times e^{-\beta V_{\rm nc}
-\frac{1}{2}\sum_{n\mu}\log{(\epsilon_{n\mu}})} \,
 e^{-\frac{\beta}{2}\sum_{n\mu}D_{n\mu}^{2}/\epsilon_{n\mu}}\nonumber \\
  \label{eq:partfinal2}
  &&\times e^{-\sum_{k=1}^{N_B}\phi^{(k)}(({\cal M}-\mu_{k})^{2}+\nu_{k}^{2})\phi^{(k)}}
\end{eqnarray}
where ${\cal M}$ is scaled as in Eq. \ref{eq:defop2} and the integrals
over $\vec{r}_i$ are taken to be sums when the charges are
constrained to be on lattice sites.  The effective Hamiltonian
appearing in the exponent terms of (\ref{eq:partfinal2}) is completely
local: to simulate the system, we need only update, in some
conveniently chosen order, (i) the locations $\vec{r}_{i}$ of the
charged entities, (ii) the lattice displacement field $D_{n\mu}$
(respecting locally the Gauss' Law constraint), and (iii) the $N_B$
auxiliary scalar fields $\phi^{(k)}$, in all cases according to the
indicated Boltzmann weight.

\section{Results}
\label{sec:results}

In order to see the effect of the additional dipole-like interaction
present in the uncorrected simulation and study the effectiveness of
our correction scheme, we have simulated a system of neutral particles
similar to the system studied in Ref.~\cite{maggs:alone}.  In
classical electrodynamics, neutral particles, even those with a
different dielectric constant from the background, do not interact and
therefore the density-density structure factor, $S(q)$, is constant as
a function of $q$.  In contrast, the simulations in
Ref.~\cite{maggs:alone} show that using the local lattice approach
(not corrected to remove the extra dipole term) to simulate neutral
particles with different dielectric constants from the background
results in a clearly non-constant density-density structure factor.
In this section we show that the multiboson method presented in this
paper is able to effectively remove the dipole terms so that the
neutral particle simulation gives a constant density-density structure
factor.  We then apply the method to the more interesting case of a
one-component plasma with charged particles of different dielectric
constant than the background.

\begin{figure}
\centerline{\includegraphics[width=3.5in]{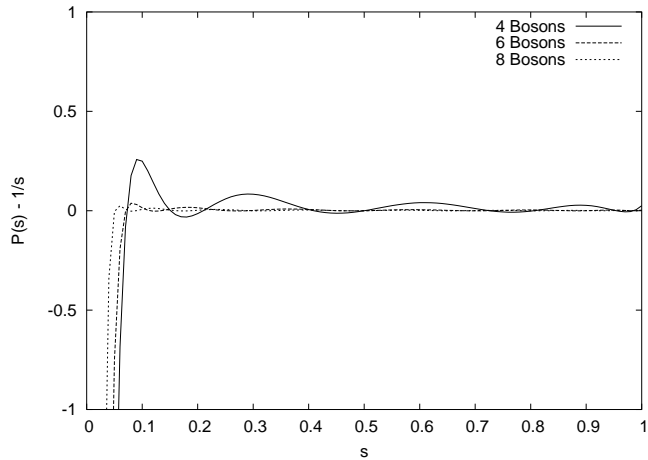}}
\caption{Residual from polynominal approximation used in the
multiboson correction factor for the simulations in this paper.}
\label{fig:polyplot}
\end{figure}

In the simulations described below we have compared the effect of
using varying number of multiboson fields to determine the sensitivity
of our results to the low eigenvalues of the operator ${\cal M}$
discussed in the preceding section. In particular, we have simulated
neutral systems with $N_B$=4 boson fields (and $\delta=$0.07) and
charged systems with $N_B=$4,6,8 (and $\delta=$0.07,0.07, and 0.05
respectively).  It is instructive to see the accuracy of the
polynominal approximation $P(s)$ to $1/s$ for these choices by plotting
the residual $P(s) - 1/s$, as shown in Fig.~\ref{fig:polyplot}.

We first consider a system of 1000 neutral particles on a $16^3$ lattice.
The particles are constrained to the lattice site and only one
particle is allowed on each lattice site.  The background dielectric
constant was set to $1.0$, and the dielectric constant of the
particles was set to $0.2$, except in the uniform dielectric
simulations where the particles had dielectric constant $1.0$.  The
dielectric constant along a link in the $\mu$ direction from a lattice
site $n$ is defined through the relation 
\begin{equation}
\frac{2}{\epsilon_{n\mu}} = \frac{1}{\epsilon_n} + \frac{1}{\epsilon_{n+\mu}},
\end{equation}
where $\epsilon_n$ and $\epsilon_{n+\mu}$ are either the background
dielectric constant or the particle dielectric constant depending on
whether there is a particle on site $n$ or site $n+\mu$ respectively.  The
dimensionless inverse temperature, $\hat{\beta} = 4 \pi e^2 /a$, was set to
$0.25$.  We performed 5000 Monte Carlo warmup sweeps, followed by
$40,000$ Monte Carlo sweeps.  To update the electric field we used the
local heat bath method discussed in Ref.~\cite{ourpaper:fft}.  To
update the neutral particles at each Monte Carlo step we chose $1000$
particles, which we attempt to move using the methods discussed in
Section~\ref{sec:howitworks}. 

\begin{figure}
\centerline{\includegraphics[width=3.5in]{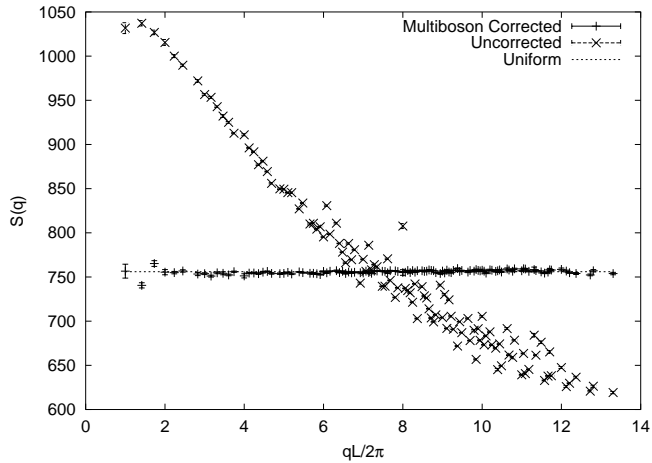}}
\caption{Density-density structure factor for the neutral system.
Shown is the results from both a simulation with the multiboson
correction factor and an uncorrected simulation.  The dotted line
gives the desired $q$ independant structure factor.}
\label{fig:neut_struct}
\end{figure}

Figure \ref{fig:neut_struct} shows the density-density structure
factor for both the corrected and uncorrected simulation.  In the
uniform case the structure factor is independent of $q$ and is given
by $N(1-\frac{N-1}{V-1})$, where $N$ is the number of particles and
$V$ is the volume of the system (the number of points in the lattice).
The spurious dipole-like interactions present in the uncorrected
simulation give a density-density structure factor that differs
significantly from the desired flat structure factor of the uniform
case, while the simulation with the multiboson correction reproduces
the analytically calculated flat structure factor of the neutral
system.  The acceptance rate for particle moves drops from 30\% in the
uncorrected simulation to 15\% in the multiboson simulation.  The
uncorrected simulation took 11 hours to run on a single processor
workstation and the corrected simulation took 55 hours on a similar
workstation.

After verifying that the multiboson method was able to correctly
remove the non-physical dipole interaction from the simulation of
neutral particles, we applied the multiboson technique to the
simulation of a one-component charged plasma.  The particles in the
system are positively charged with one electron charge each and the
background sites are uniformly given enough negative charge so that
the entire system is charge neutral.  The parameters of the system are
the same as in the neutral particle case except the lattice is a
$32^3$ lattice, there are $8000$ particles, $8000$ particles are
updated each Monte Carlo step, and the dielectric constant of the
particles in the non-uniform case was taken to be $0.05$. In this case
we have also repeated the simulations with varying numbers of boson
fields ($N_B$=4,6,8).

\begin{figure}
\centerline{\includegraphics[width=3.5in]{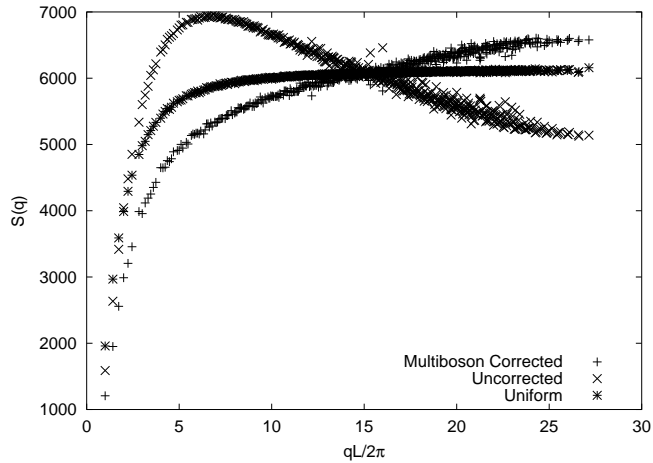}}
\caption{Density-density structure factor for the charged particles in
the one-component plasma.  Shown are the results from the uncorrected
simulation, the simulation with the multiboson correction factor, and
a simulation where the dielectric constant is uniform.  The errorbars
are smaller then the symbol size.  }
\label{fig:ocp_struct}
\end{figure}

In Fig.~\ref{fig:ocp_struct} we show the results for the
density-density structure function for three cases: the fully
multiboson corrected structure function for this system, the
uncorrected structure function switching off multiboson contributions,
and for a charged plasma with uniform dielectric ($\epsilon_{\rm part}
= \epsilon_{\rm bg} = 1.0$). It is apparent from the figure that the
removal of the spurious interactions induced by the transverse part of
the field in the nonuniform dielectric case results in a qualitative
modification of the shape of the structure function. Also, the results
differ significantly between the cases of uniform and nonuniform
dielectric.  This difference may play an important role in the
behavior of systems from biological and chemical physics, so it is
important to be able to reliably calculate the effect of having a
non-uniform dielectric {\em which changes dynamically in the course of
the simulation}. The results in Fig.~\ref{fig:ocp_struct} were
obtained using $N_B$=4 multiboson fields to estimate the transverse
electric field determinant.

\begin{figure}
\centerline{\includegraphics[width=3.5in]{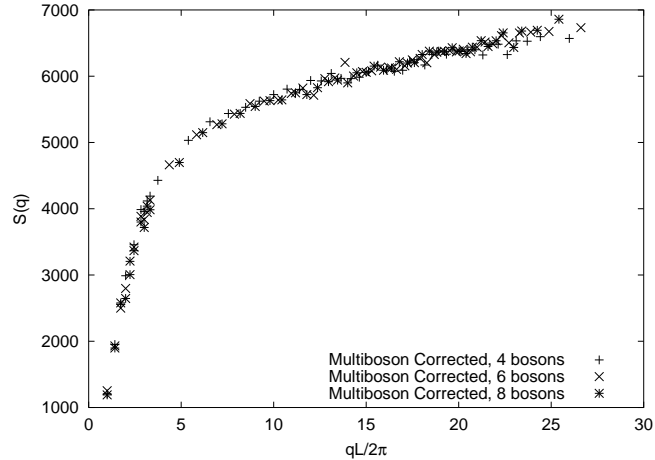}}
\caption{Density-density structure factor for the charged particles in
the one-component plasma.  Shown are the results from simulations
corrected with differing numbers of bosonic field in the multiboson
correction factor.  The errorbars are smaller then the symbol size.  Only a portion of the larger $q$ results are shown to improve the legibility of the plot.}
\label{fig:ocp_struct_all}
\end{figure}

Of course, the polynominal approximation Eq.\ \ref{eq:chebpoly} cannot
accurately represent very small eigenvalues of ${\cal M}$, so it is
important to check the sensitivity of our results to the number of
boson fields used. This comparison, again for the case of the charged
one-component plasma on a 32$^3$ lattice described above, is shown in
Fig.4. Certainly the results for $N_B$=4,6, and 8 are in complete
qualitative agreement over the entire range of $q$. For larger lattices,
one will need to increase $N_B$ to deal with the larger dynamical range
in the eigenspectrum of ${\cal M}$, and, as we saw previously, this in
turn results in a drop in the acceptance rate for particle moves. A
similar problem in the use of multiboson fields in unquenched QCD has
been addressed \cite{qcd-multiboson} by use of a hybrid scheme in which
the number of multiboson fields is held fixed, and the low eigenvalues of
${\cal M}$ treated exactly by a Lanczos algorithm.

\section{Conclusion}
\label{sec:conclusion}

Systems from biological and chemical physics frequently have mobile
charged elements with different dielectric constant from the
background.  The local lattice approach provides an efficient method
to simulation these systems.  Unfortunately, a spurious force remains
as an artifact of the simulation method.  This force can be
efficiently removed using a set of bosonic fields to approximate a
non-local counteracting force.  This method is able to effectively
remove the spurious term in both neutral and charged systems.

\begin{acknowledgments}

The work of A.~Duncan was
supported in part by NSF grant PHY0244599. A.~Duncan is grateful
for the hospitality of Max Planck Institute (Heisenberg Institut f\"ur 
Physik und Astrophysik, Munich), where part of this work was performed.
The work of R.D.~Sedgewick
and R.D.~Coalson was supported by NSF grant CHE0092285.
\end{acknowledgments}


\begin{thebibliography}{99}
\bibitem{maggs:prl} A.C.~Maggs and V.~Rossetto, Phys.\ Rev.\ Lett.\ \textbf{88}
, 196402 (2002)
\bibitem{maggs:alone} A.C.~Maggs, J.\ Chem.\ Phys.\ \textbf{120}, 3108
  (2004)
\bibitem{maggs:worm} L.~Levrel, F.~Alet, J.~Rottler, and
A.C.~Maggs, Statphys22 Proceedings, to be published in PRAMANA
[also available at  \texttt{cond-mat/0409350}]
\bibitem{maggs:trail} L.~Levrel and A.C.~Maggs, \texttt{cond-mat/0503744} (2005)
\bibitem{ourpaper:fft} A.~Duncan, R.D.~Sedgewick, and R.D.~Coalson,
  \texttt{cond-mat/0508266} (2005)
\bibitem{ourpaper} A.~Duncan, R.D.~Sedgewick, and R.D.~Coalson, Phys.\ Rev.\ E \textbf{71}, 046702 (2005)
\bibitem{sitkoff} D.~Sitkoff, K.A.~Sharp, and B.~Honig, J.\ Chem.\
  Phys.\ \textbf{98}, 1978 (1994)
\bibitem{kessel} A.~Kessel, D.S.~Cafiso, and N.~Ben-Tal, Biophysical
  Journal \textbf{78}, 571 (2000)
\bibitem{luescher} B. Bunk, K. Jansen, B. Jegerlehner, M. L\"uscher,
  H. Simma, and R. Sommer, \texttt{hep-lat/9411016} (1994)
\bibitem{alet:orig_worm} F.~Alet and E.~S\o rensen, Phys.\ Rev.\ E
  \textbf{67}, 015701 (2003)
\bibitem{ewalda} J.V.L.~Beckers, C.P.~Lowe and S.W.~de~Leeuw,
  Molecular Simulation \textbf{20}, 269 (1988)
\bibitem{ewaldb} J.W.~Perram, H.G.~Petersen and S.W.~de~Leeuw,
  Molecular Phys.\ \textbf{65}, 875 (1985)
\bibitem{QCD:unquenched} I.~Montvay and G.~M\"unster, \textit{Quantum
  Fields on a Lattice}, (Cambridge University, Cambridge, 1994)
\bibitem{fft} E.~Esssmann, L.~Perera, M.L.~Berkowitz, T.~Darden,
  H.~Lee and L.G.~Pedersen, J.\ Chem.\ Phys.\ \textbf{103}, 8577 (1995)
\bibitem{tonyqcd} A.~Duncan, E.~Eichten and H.~Thacker,
  Phys.\ Rev.\ Lett.\ \textbf{76}, 3894 (1996)
\bibitem{qcd-multiboson} A.~Duncan, E.~Eichten and J.~Yoo, Phys.\ Rev.\ D \textbf{68},
 054505 (2003)
\end{thebibliography}
\end{document}